# Development of a Practicable Digital Pulse Read-out for Dark-field STEM

Tiarnan Mullarkey[1,2,*], Clive Downing[3], Lewys Jones[1,3]

1. *School of Physics, Trinity College Dublin, Dublin 2, Ireland*
2. *Centre for Doctoral Training in the Advanced Characterisation of Materials, AMBER Centre, Dublin 2, Ireland*
3. *Advanced Microscopy Laboratory, Centre for Research on Adaptive Nanostructures and Nanodevices (CRANN), Dublin 2, Ireland*

**Abstract**

When characterising beam-sensitive materials in the scanning transmission electron microscope (STEM), low-dose techniques are essential for the reliable observation of samples in their true state. A simple route to minimise both the total electron-dose and the dose-rate is to reduce the electron beam-current and/or raster the probe at higher speeds. At the limit of these settings, and with current detectors, the resulting images suffer from unacceptable artefacts including; signal-streaking, detector-afterglow, and poor signal-to-noise ratios (SNR). In this manuscript we present an alternative approach to capture dark-field STEM images by pulse-counting individual electrons as they are scattered to the annular dark-field (ADF) detector. Digital images formed in this way are immune from analogue artefacts of streaking or afterglow and allow clean, high-SNR images to be obtained even at low beam-currents. We present results from both a ThermoFisher FEI Titan G2 operated at 300kV and a Nion UltraSTEM200 operated at 200kV, and compare the images to conventional analogue recordings. ADF data are compared with analogue counterparts for each instrument, a digital detector-response scan is performed on the Titan, and the overall rastering efficiency is evaluated for various scanning parameters.

[*] Corresponding author: *mullarkt@tcd.ie*

**Introduction**
In the scanning transmission electron microscope (STEM) a finely focused electron probe is rastered across the sample surface generating a multitude of spatially resolved structural and chemical signals. One of the most frequently adopted modes is annular dark-field imaging (ADF-STEM). This mode is so widely used because its incoherent image formation mechanism gives readily interpretable mass/thickness contrast, sometimes referred to as z-contrast (Krivanek et al., 2010; Jones, 2016).

As STEM imaging studies become increasingly rigorous and quantitative, the risk of possible electron-beam induced damage arises. When presenting the results of, for example, a prolonged chemical mapping, a diligent study will often present a 'before and after' image pair to assess samples damage and to infer the reliability of the data captured in between.

The electron-dose falling on a sample is measured in electrons-per-unit-area. For high resolution imaging this would usually be in electrons per square Angstrom ($e^-/Å^2$). Under typical conditions it is not uncommon to use ~$10^8$ $e^-/Å^2$ while imaging stable samples (Gnanasekaran et al., 2018), however this is maybe a million times too high for fragile samples (Buban et al., 2010). For example, zeolites and organic single crystals show damage starting from the order of 100 $e^-/Å^2$, with some examples of the latter damaging at doses far less than this (S'ari et al., 2019; Revol & Manley, 1986; Pan & Crozier, 1993). Biological materials are notoriously sensitive to beam-damage, and are easily damaged at doses of 10 $e^-/Å^2$ and below, even when stabilised by using methods such as cryo-EM (Frank, 2002; Glaeser, 1971).

For beam-robust specimens, and at high electron-doses, it is readily possible to reach the ultimate *instrumental resolution* of a modern aberration corrected STEM (often well below 1 Å). However, for beam-sensitive samples, where it is not possible to use high beam-currents, the *dose-limited resolution* becomes the limiting performance (Egerton et al., 2004; Egerton, 2014). Moving beyond image resolution, and considering information *precision*, we again see a dependence on dose (De Backer et al., 2015).

While the term low-dose imaging can vary in definition, by maybe three or five orders of magnitude between say structural biology or semiconductor metrology, there is one ethos that should be universal; that however many electrons must be used should be detected in the most efficient way. The coming generation of pixelated STEM detectors offer an exciting and potentially very efficient approach here, especially for light elements (Pennycook et al., 2015; Yang et al., 2016); however, at present they are around 100x slower than conventional STEM detectors, 100x more expensive, and perhaps 100x less prevalent within microscopy centres. Here instead we examine a route to extend the performance of the current generation of ADF detectors already equipped on instruments in place of these potentially expensive upgrades.

This manuscript is structured as follows; we first introduce a new practicable method of digital imaging by electron-counting, and the effects it has on images and quantitative-ADF calibration, including demonstrations with real experimental data from a variety of instruments. Next, we show the potential for improvements in dynamic-range of more than 600x compared with previous similar attempts. Finally, the effect of long flyback times relative to ever shorter dwell-times is discussed and the implications for future experiment design for low-dose acquisition.

**Background**
For many studies, it becomes clear that dose-management becomes crucial in ensuring maximum overall performance. However, lively debate still exists about the precise sample-damage roles of *dose* versus *dose-rate* (Jiang & Spence, 2012; Johnston-Peck et al., 2016), and in the special case of multi-frame STEM spectrum imaging (where beam-current and number-of-frames offer new independent experiment design parameters) the *instantaneous-dose* into one pixel even appears to be relevant even when both overall dose and dose-rate remain unchanged (Jones et al., 2018). What all investigators can agree on however, is that whichever metric is chosen that lower is better if the goal is to minimise sample damage.

The electron-dose per unit area received by a sample is given by Equation 1:

$$Dose = \frac{I \cdot C \cdot \delta_t}{(dx)^2} \quad (1)$$

Where $I$ is the probe current, $C$ is the Coulomb number, $\delta_t$ is the pixel dwell-time in seconds, and $dx$ is the pixel-width.

Alternatively, moving away from conventional (Shannon) scanning, compressed sensing (CS) has been suggested as a route to reduce sample beam-exposure (Stevens et al., 2014; Kovarik et al., 2016). In this special case of CS Equation 1 is modified by a constant multiplying factor:

$$Dose = \frac{I \cdot C \cdot \delta_t}{(dx)^2} \cdot F_C \quad (2)$$

where $F_c$ is a number less than one describing the fraction of pixels un-blanked in the CS scan. However, depending on the CS implementation this requires additional costly beam-blanking or rastering hardware. Moreover, for systems limited by purely Poisson noise, CS has not been shown to deliver additional information over conventional scanning (Sanders & Dwyer, 2020; Van den Broek et al., 2019).

From Equation 1 we can see that in a conventional scanning experiment, low-dose conditions can be achieved by either enlarging the pixel-size, reducing the beam-current or by reducing the pixel dwell-time. When seeking to resolve some given feature, pixel size may not exceed Nyquist sampling of that size, so arbitrarily enlarging this is not an option (Shannon, 1949). Further, there are often practical limits in the instrumentation about how low the emission can be reduced and remain stable, so increasing scan-speed initially appears an appealing route. However, Buban et al. demonstrated some of the limitations to simple scanning faster such as image streaking (Buban et al., 2010). As shown later, streaking can originate from both the response of the detector to single electrons and afterglow of the scintillating crystal in the detector.

At the extremes of low beam-current and fast dwell-times, the efficiency, noise-behaviour, and response-speed of the ADF detector then become key parts of the image formation system, quickly become what limit the data-quality and interpretability. Thus, for very beam-sensitive materials such as nano-particles or zeolites, simply stretching these parameters is not sufficient, and truly low-dose (or low dose-rate) imaging strategies must be developed.

When operating at very low beam-currents, it has been seen that the normally recorded continuum ADF intensity becomes discretised, with apparently single electron events becoming visible (Ishikawa et al., 2014; Krause et al., 2016; Sang & LeBeau, 2016; Mittelberger et al., 2018). Previously, Krause et al. (Krause et al., 2016) and Sang & LeBeau (Sang & LeBeau, 2016) recorded oscilloscope traces from ThermoFisher Titan (S)TEM instruments fitted with Fischione ADF detectors and found the profile of each event to be well described by a sharp onset followed by an exponential decay; this is also observed in other photomultiplier tube (PMT) characterisation studies (Deng et al., 2013). Mittelberger et al. recorded ADF images with short dwell-times, using the in-built scan-unit on a Nion dedicated STEM, and observed a Lorentzian shape (Mittelberger et al., 2018).

The concept of electron-count imaging is not new to electron microscopy. Studies from as early as the 90s noted a decrease in noise when imaging using electron counting in SEM. A comparative study demonstrated that electron count images were of higher quality with respect to SNR, contrast, and resolution when compared to conventional analogue detection (Uchikawa et al., 1992; Yamada et al., 1991). In this work, we seek to realise these same improvements in the scanning TEM.

**Methods**
*Experimental Setup*
A portable TektronixTBS1022 oscilloscope was used to determine the correct pins/connectors for monitoring scan-generator voltages and the ADF output. It should be noted that PMTs include kV power supplies so this should only be done under the supervision of a competent person.

Previous studies have shown the response time of common ADF PMTs to be of the order of 1-3μs (Krause et al., 2016; Sang & LeBeau, 2016; Mittelberger et al., 2018). To accurately distinguish and time the arrival of a single electron pulse we would need to sample this at least 4 times across the peak. A PicoScope 2206B USB streaming oscilloscope/DAQ was used for data capture. It was chosen for its sampling rate, ability to stream data to MATLAB, and its relatively low cost. It has a sampling rate of >32MHz and the data buffer is only limited by the host PC's RAM. This sampling rate is sufficient to accurately capture pulses and is more than 15 times faster than the Gatan Digiscan system (2MHz clock).

When recording data, the gain of the PMT is set at the microscope to avoid clipping (either high or low) of individual electron pulses. Next, suitable voltage ranges and offsets were selected for the DAQ which maximised use of its voltage sampling bit-depth while again avoiding clipping.

The image size and pixel dwell-time are also set at the microscope and are noted for later reshaping of the data stream into an image. Other parameters necessary to do this are the line flyback time ($T_{LFB}$), and the time taken for the beam to travel from the end of one image frame to the beginning of another, herein referred to as the frame flyback time ($T_{FFB}$). On the Nion used, here controlled via a Digiscan, the $T_{LFB}$ can be set by the user. Whereas when using the Titan (with its OEM scan-gen) it cannot. Instead it was measured using the Tektronix oscilloscope to probe the appropriate clock signals. This is also how the $T_{FFB}$ was measured on both instruments.

Having interfaced the PicoScope to MATLAB, a script collects data for a pre-set amount of time; this time is set slightly longer than needed for a complete number of scan frames. Once captured the data is digitised and reshaped into an image.

*Digitisation Process*
When operating at ultra-low beam-currents, individual scattered electrons cause discrete voltage pulses at the PMT that are distinguishable from one another. An example of this raw-trace is shown in Figure 1 (top).

This typical data shows both individual pulses, 'A', and compound (pile-up) pulses, 'B'. In previous work these compound pulses were not separable using simple thresholding (Mittelberger et al., 2018). However, each pulse has a sharp rising edge (large positive gradient), a peak (zero gradient), and a decaying edge (smaller negative gradient). As such, the *gradient* of the analogue output (Figure 1, middle) contains abrupt peaks centred on the rising edge when each electron hits the detector.

Electron impacts can then be readily distinguished by applying MATLAB's *findpeaks* function to this gradient for more sensitively than they could be by applying the same to the recorded voltage signal.

The user can input a threshold into the *findpeaks* function which defines the minimum height difference between a data point and its neighbours for it to be recognised as a peak. A secondary script was created which displays a section of the gradient as well as a histogram of the gradient values present in the data. This allows an appropriate threshold to be chosen.

Note how the previously compound pulse in Figure 1 is clearly resolved into four separate pulses. Hence, a digital signal is created where each '1' is a single electron impact and otherwise '0', Figure 1 (bottom).

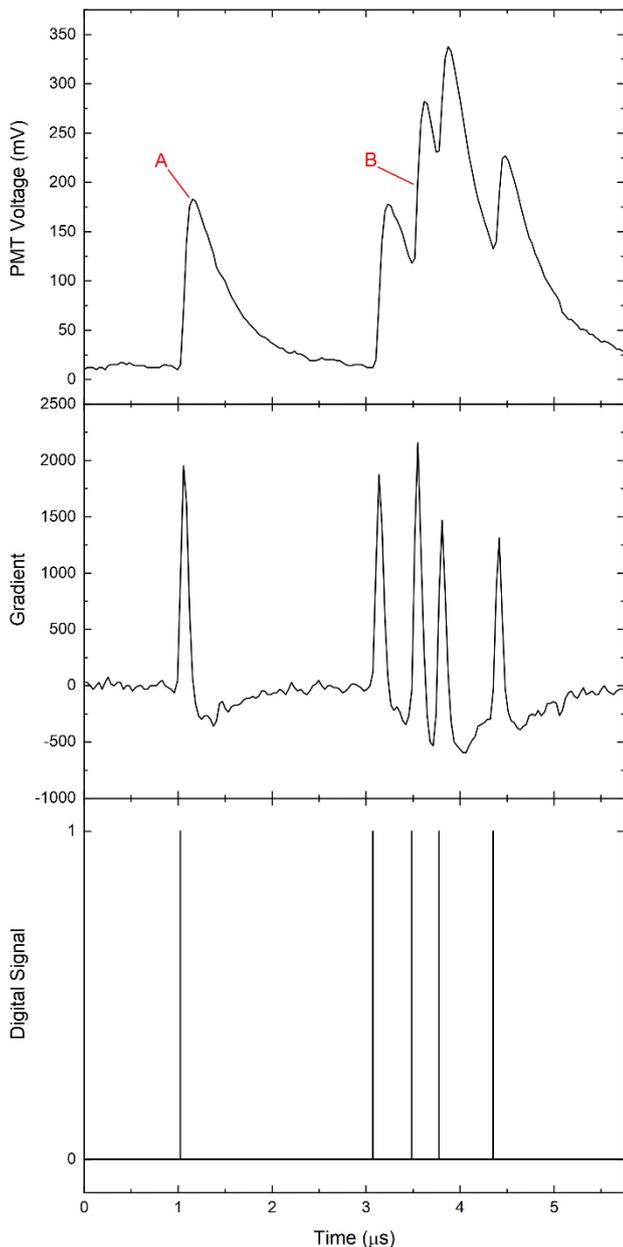

*Figure 1. A nearly 6μs span read-out of an ADF detector with 5 electron impacts visible taken with a sampling rate of 31 million samples per second. The raw analogue signal, gradient, and digital signal are shown.*

This digital signal is still a one-dimensional data stream and needs to be reshaped to yield an image. First the binary signal is integrated (binned) to form pixels where the binning ratio is merely the ratio of the sampling time and the dwell-time. Next, using parameters from the clock-signals ($T_{LFB}$, $T_{FFB}$ image-dimensions etc.), these pixels are combined into lines, and the lines into images.

As the PicoScope is retrofitted to existing equipment, and merely observes the scan rather than controlling it, a partial frame is often captured before the full scan-frame(s). To accommodate this, slightly more data is captured than is needed and the excess can be discarded later. The scattering during the line-flyback is also observed and is discarded when presenting the image.

*Samples*

Although eventually the interest lies with beam-sensitive specimens, to verify our approach we begin by imaging a silicon lamella and a sample of gold nanoparticles on a support of amorphous carbon. This allows us to verify the precision of the data-to-frame reshaping with the lamella, and the improvement to dynamic range with the mass range on the supported nanoparticles.

*Experiment Design & the Maximum Dose-rate*

Pulse read-out requires that as many as possible of the individual electron detection events be separately identifiable. This imparts an upper limit to the detection rate before events pile up and cannot be separated (similar to dead-time in x-ray detectors).

For a detector-PMT-amplifier system with an afterglow half-life of ≈ 2.5μs, to readily discriminate between sequential impact events, it is preferable to have a maximum event frequency on average of one every 5μs (=200kHz, or one per 10 pixels of 0.5μs dwell-time). For ADF, imaging where perhaps up to 10% of primary electrons are scattered to the annular detector, this is equivalent to $2 \times 10^6$ primary electrons per second or ≈ 0.32 pA. Even using a relatively fine pixel-size of 0.1 Å, this equates to a *maximum* electron dose for this technique of 100 $e^-/Å^2$, which would not be expected to deliver sufficient precision for many physical science studies (Van Aert et al., 2019; De Backer et al., 2015).

The above was considering being able to distinguish pulses in the detector output without first taking the gradient. However, by doing so we see that peaks in the gradient corresponding to electron impacts are ~10 times narrower, Figure 1 (middle), and are not required to be separated by so many fallow pixels. This results in the ability to digitise far more electrons, around 30x more (3,000 $e^-/Å^2$), scattered to the detector. This is of particular importance for samples containing both light and heavy elements which strongly vary in scattering strength.

Still further, where two or three annular detectors are used simultaneously, this increases the maximum combined event rate to 18,000kHz (18 MHz) or ≈ 9,000 e$^-$/Å$^2$, which brings opens up an optimal dose range for some materials science studies (Van Aert et al., 2019).

Where specialised multi-ring/segment detector-PMT combinations are used, such as a 16 segment detector (Shibata et al., 2010), this expands the maximum dose-rate envelope perhaps to 48,000 e$^-$/Å$^2$ and offers the potential for greatly reduced dose differential phase contrast imaging (DPC) (Shibata et al., 2012). While such configurations may not offer the full flexibility of a pixelated STEM detector (Pennycook et al., 2015), the reduced data burden and far faster frame-rates achievable using PMT based systems make TV-rate imaging of dynamic events possible.

Additionally, because the pulse read-out signal is fully digital the initial setup and calibration of dark and gain settings in such configurations can be greatly simplified and experiments streamlined.

## Results and Discussion

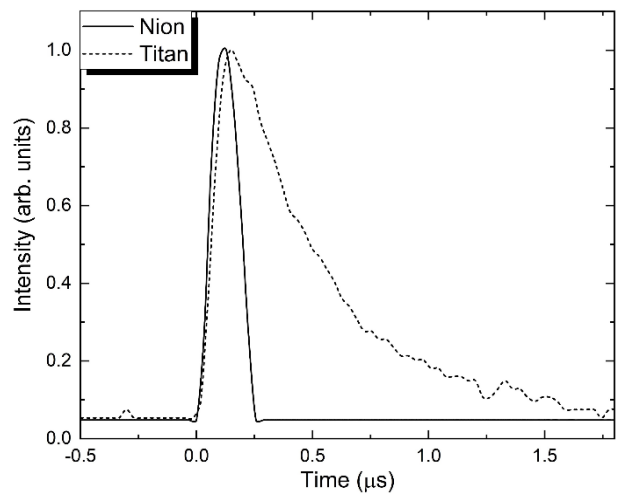

*Figure 3. Individual electron pulse responses at the detector recorded with the maximum sampling rate of the PicoScope, 31.25MHz.*

### *Electron Pulses and Digitisation*

To ensure the generality of the approach, both a ThermoFisher Titan-G2 80-300 and a Nion UltraSTEM200 were studied.

Having determined a beam current which results in little to no overlap between consecutive pulses, this was then used to capture clean single pulse profiles (Figure 3).

Although the shape of the pulse due to an electron impact varies by instrument (exponential decay or Lorentzian), each exhibits a sharp rising edge and can still be digitised by the same approach. It was even found that on the Nion the width of the pulse

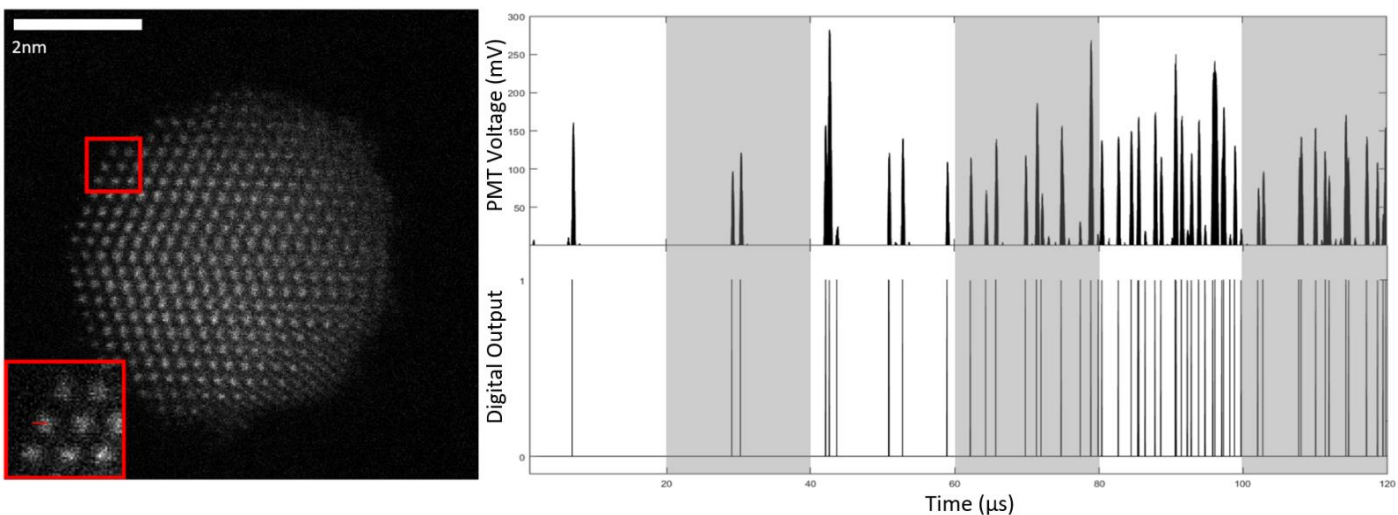

*Figure 2. Experimental digital image of gold atoms on an amorphous carbon background. Captured with a 20μs dwell-time. The line profile (indicated in red) demonstrates digitisation of signal from both carbon and gold and extends across six pixels, which alternate grey and white.*

depends on where the data is streamed from, but both shapes work with our approach (Figure S1).

When imaging under these ultra-low-dose conditions a conventional analogue ADF image is dominated by pixels containing no scattered electrons (Sang & LeBeau, 2016). These dark-pixels contain only noise in the form of the D.C offset and its associated dark-noise. Superimposed on this background are electron impact events of varying intensity, where the intensity depends on the location of the ADF detector where the scattered electron impacts (Macarthur et al., 2014; Krause et al., 2016).

By its nature all electron impacts in the digital signal are localised to a single sampling interval (and hence a single image pixel), and are also detected with equal efficiency, this satisfies the definition of perfect modulation transfer function (MTF) and detector quantum efficiency (DQE) (McMullan et al., 2009).

*Reduction of Imaging Artefacts*
In all images shown in this manuscript, the fast-scan direction is horizontal and from left to right with the scanning beginning in the top left of the frame.

The decay time of electron impacts in the analogue image is of the order of a few microseconds, and so when scanning at short dwell-times (<2us), the signal of an impact can spread across multiple pixels, potentially resulting in the loss of high-resolution information. As well as this, the digital signal records only electron impacts as signal, and so any D.C offset and background noise is absent in the digitised image. To clearly visualise these effects a low-magnification image of a silicon lamella was captured.

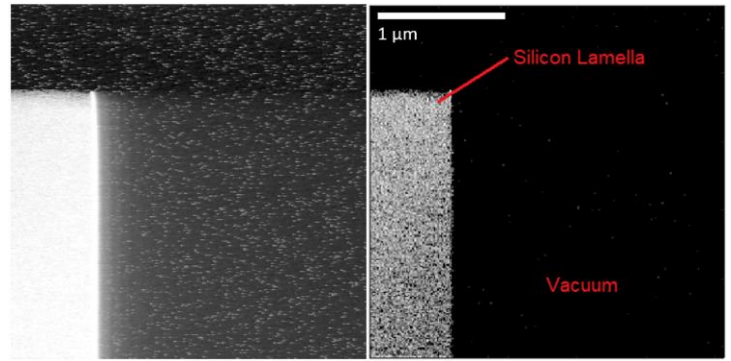

*Figure 5. Example analogue (left) and digital (right) images of a silicon lamella. The image gamma has been exaggerated to reveal streaking from both individual electron signals and detector afterglow.*

There are two features to note in the above image comparison. Firstly, the appearance of the vacuum. For a dark-field image, the vacuum region should show no scattering to the detector and appear black. In the analogue image the presence of background noise and detector afterglow result in a vacuum which is far from black, but in the digital image the vacuum is perfectly dark, excluding thermally generated electrons. Secondly, the streaking of individual pixels can be seen in the analogue image, but this is absent in the digital image. The combination of these effects results in digital images with a higher signal-to-noise ratio when imaging under these conditions.

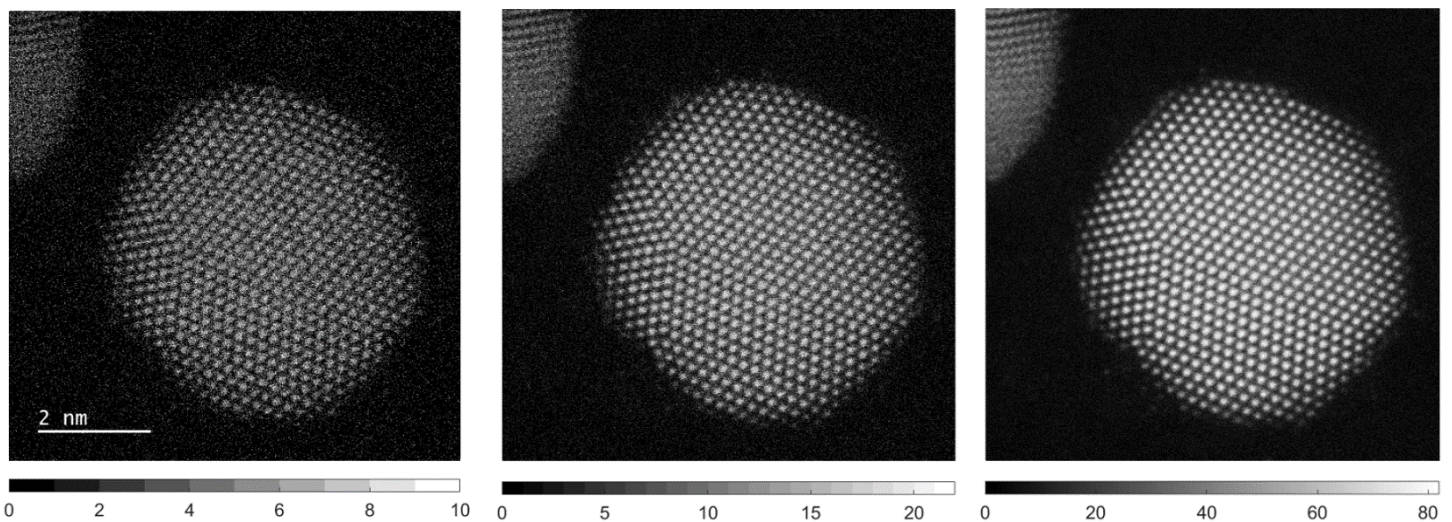

*Figure 4. One frame, four frames, and 20 frames which have been rigidly aligned and summed. Each frame was captured with a 2μs dwell-time and ~5pA beam-current. The colourbar has units of integer number of electron impacts per pixel. Notice the discretised greyshades for the single frame become a continuum as more frames are summed.*

Previous attempts in the literature to perform electron counting imaging have struggled where both strong and weak scatterers were present in a single field-of-view such as gold nanoparticles on amorphous carbon (Mittelberger et al., 2018). In that work, strong scattering from gold columns caused the rate of electron being scattered to the detector to overwhelm the 6MHz clock of the Nion Superscan (with a Gatan Digiscan this situation would have been even worse). However, by using the 32MHz DAQ and the gradient-based approach presented here, this issue can be avoided and it is possible to produce a digital image on gold atoms (Figure 3).

The line profile in *Figure 3* begins on the carbon support and extends across a gold atom, this is noticeable as the increase in rate of electron impacts seen in the PMT output. The ability for the oscilloscope used to record multiple data-points per image pixel and the gradient-based approach to digitisation is what allowed this image to be captured where previous attempts have not been successful.

*Expanding Dynamic Range for Digital ADF*
Previously Mittelberger et al. devised a method of electron-count imaging for use with a STEM. It was reported that for their method to work that the "signal level has to be well below 1 e/px on average". As such, although they were able to image carbon with this method, they were unable to image gold atoms as they scattered electrons too frequently due to their larger atomic mass.

The method presented in this paper suffers from no such restriction, and we successfully produced images of gold atoms on carbon (figures 4, 5). This was possible due to both the increased sampling rate of the PicoScope and the gradient-based approach to digitisation. This method has multiple readouts per image pixel so we are not limited by the scan speed of the scan generator used. Furthermore, peaks due to electron impacts are sharper and easier to distinguish in gradient-space than they are in real-space. This overall increase in temporal-resolution is the key to this method and allows the benefits of digital imaging to apply to images containing both light and heavy elements. The benefits beyond a higher SNR are discussed below.

The approach of binning multiple fast samples into each pixel dwell-time already increases the potential dynamic range by a factor of up to 30 for a 2us dwell-time (with this factor increasing with longer pixel dwell-times). While this dynamic range is already sufficient to yield atomic resolution imaging (Figure 3), it can be further improved by summing multiple aligned scan frames (Jones et al., 2015). To demonstrate this, multiple digital frames of a fresh region of the Au nanoparticle sample were captured with a 2μs dwell-time and ~5pA beam current. These frames were then rigidly aligned and summed, resulting in an increase in dynamic range of ~ 600x (Figure 5).

*Beam-normalised Quantitative ADF*
For quantitative ADF scattering studies, often the starting point in the data analysis is to subtract (or fit to) some constant background (De Backer et al., 2016), or to subtract some smoothly varying ramp representing the thin support (De Backer et al., 2015). However, in the presence of D.C. vacuum noise (which we can see here in the analogue data in Figure 1-top, and Figure 5-left), this choice of background subtraction can itself be a source of error. Often the minimum pixel-value is taken, or the average from some vacuum area. However, with the proposed digital read-out an unambiguous zero baseline is directly established and no further D.C. subtraction or fitting step is needed. This in turn maximises the reliable contrast-to-noise ratio (CNR) which can lead to more reliable peak-finding in atomic-resolution data (Fatermans et al., 2019).

For further comparison both an analogue and digital detector sensitivity scan were taken of the Titan's Fischione model 3000 ADF detector, both shown in Figure 6.

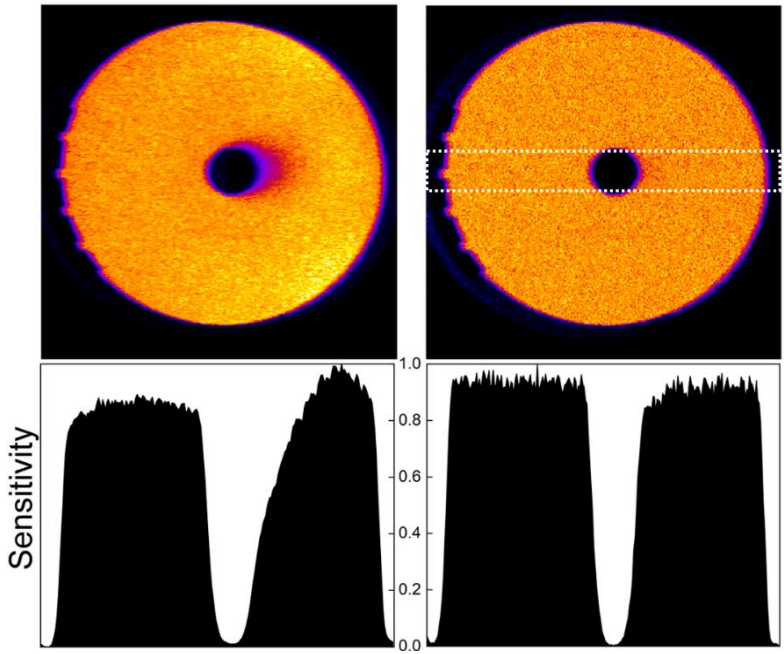

*Figure 6. Comparison of conventional (analogue) Fischione model 3000 ADF detector sensitivity scan (left), and the digital equivalent (right). The digital read-out shows a flatter and more homogeneous response.*

The sensitivity profile in the above image is the average of 30 line profiles from the boxed region normalised by the largest value of each scan.

In a perfect ADF recording system, the signal of each scattered electron should be equal regardless of the angle it was scattered through. However, less sensitive parts of the detector surface lead to a lower output and appear as dark regions in the analogue detector scan (Findlay & LeBeau, 2013; Macarthur et al., 2014). In the digital signal all electrons are recorded with equal intensity regardless of their intensity at the PMT output, and so these dark regions are absent. Having a more homogeneous detector response such as in the digital scan also facilitates easier inner-angle measurements.

*Dose Efficiency*
As the streaming-oscilloscope is continually recording data this provides the opportunity to observe the behaviour of the beam during the flyback time. This relatively short time is usually hidden from the operator and the time and dose therein is often ignored. The images presented so far have been trimmed to show the same view as would be shown to the user sitting at the instrument. Please refer to supplementary image S2 to see an example of an image from which the flyback has not been trimmed.

Seeing this image raises questions about the dose a sample being imaged truly receives, which we discuss in this section.

Due to this behaviour of the beam during the line flyback time, the previously introduced dose equation (1) does not accurately calculate either the electron dose received by the sample within the frame of view, nor the sample as a whole.

This was evidenced when reshaping the lamella where it was seen that not only did the beam travel back across the part of the sample being imaged, but even continued and hit parts of the sample outside the field-of-view. Thus, the dose equation does not accurately reflect the electron-dose the region of interest receives due to the double-scanning, but also does not account for damage to other parts of the sample.

Instead a new measure of efficiency, η, is proposed. η is the ratio of useful, information collecting time, to the duration of an entire frame time, which represents the time the sample may potentially suffer beam damage.

$$\eta = \frac{Useful\ Information\ Time}{Total\ Exposure\ Time} \quad (3)$$

$$\eta = \frac{\delta_t * n_P}{(\delta_t * n_P) + (T_{LFB} * n_L) + T_{FFB}} \quad (4)$$

Where $n_P$ is the number if image pixels and $n_L$ is the number of scan-lines in the image, and the other terms are as previously defined.

The denominator in this expression can either be measured experimentally, for example simply timing ten or twenty frames with a stopwatch, or it may be calculated explicitly from the timing data on each line and between frames (Figure S3). Some typical results for the conditions used are shown in Table 1.

*Table 1. Calculated dose efficiencies for some of the conditions used in the results shown in this manuscript.*

|  | $\delta_t$ | Flyback Time | Image Size | EOFT | $\eta$ |
|---|---|---|---|---|---|
| **Titan** (Figure 3) | 0.5 µs | 193 µs | 256 x 256 | 321 µs | 0.397 |
| **Nion** (typical) | 2 µs | 200 µs | 256 x 256 | 110 µs | 0.719 |
| **Nion** (at min $\delta_t$ e.g. as in (Mittelberger et al., 2018)) | 0.167 µs | 200 µs | 256 x 256 | 110 µs | 0.176 |
| **Nion** (Figure 3) | 20 µs | 200 µs | 512 x 512 | 110 µs | 0.981 |

In our initial experiments, where we were trying to reproduce the work of Mittelberger et al (Mittelberger et al., 2018), a dwell-time of 0.167µs was used (minimum for Nion scan-gen's 6MHz clock) resulting in a dose efficiency of less than 18%. The dwell-time was necessarily short so as to be able to capture pulses with a high enough sampling frequency. However, by decoupling the concept of dwell-time and sampling frequency using the PicoScope it became possible to use more conventional dwell-times and for Figure 5 the dose-efficiency was increased to over 98%.

Ideally we would like to use a short as possible dwell-time with a low beam current to minimise both the electron-dose and dose-rate, however this it at odds with the idea of increasing the dwell-time to increase η. To increase η while still using a low dwell-time it can be seen in equation (4) that most appealing option is to reduce the line flyback time. Lowering the line flyback introduces new issues due to hysteresis in the scan coils, but we are currently preparing a manuscript which deals with this topic.

## Conclusions

In this work we have presented a novel approach to utilise the intrinsic sensitivity of existing scintillator-photomultiplier based STEM detectors to single electrons. By retrofitting new electronics that are more than an order of magnitude faster than a current Digiscan system, we are able to realise a fully digital ADF imaging mode. Being a digital imaging mode, all electrons are attributed both discretely to a single pixel and also with equal intensity. This eliminates both afterglow and image streaking effects, as well as quantification artefacts arising from ADF detector inhomogeneity.

The continuous data spooling nature of our prototype apparatus unfortunately laid bare the presence of significant flyback time on each scan-line and between sequential scan-frames. This time overhead leads to a significant electron-dose being deposited into the sample for no information gain. By decoupling the concept of image-pixel dwell-time and read-out sampling rate, we are able to retain our single electron sensitivity while still using conventional scanning speeds, and as such are able to increase electron dose-efficiency from less than 18% to over 98%.

The future outlook for this approach may include a shift to faster clock signal readouts, as well as a move to the new generation of solid-state diode detectors (Si-PMTs) (Buzhan et al., 2003). We would hope to make further use of other timing signals to more seamlessly automate the line and frame trimming, and in future, our approach could be coupled with novel beam-blanking schemes to eliminate entirely the wasted dose lost during flyback and between frames (Béché et al., 2016). We could also move towards in-hardware pulse-counting, however this would also lead to increased costs.

## Acknowledgments

The authors acknowledge the facilities at Advanced Microscopy Laboratory (CRANN). Tiarnan Mullarkey acknowledges the Summer Undergraduate Research Experience (SURE) programme of the School of Physics at Trinity College Dublin and the SFI Centre for Doctoral Training in the Advanced Characterisation of Materials (award reference 18/EPSRC-CDT-3581). Lewys Jones is supported by SFI/Royal Society Fellowship URF/RI/191637. The authors acknowledge Jakob Spiegelberg for helpful discussions.

## References

VAN AERT, S., DE BACKER, A., JONES, L., MARTINEZ, G. T., BÉCHÉ, A. & NELLIST, P. D. (2019). Control of Knock-On Damage for 3D Atomic Scale Quantification of Nanostructures: Making Every Electron Count in Scanning Transmission Electron Microscopy. *Physical Review Letters* **122**, 66101.

DE BACKER, A., VAN DEN BOS, K. H. W., VAN DEN BROEK, W., SIJBERS, J. & VAN AERT, S. (2016). StatSTEM: An efficient approach for accurate and precise model-based quantification of atomic resolution electron microscopy images. *Ultramicroscopy* **171**, 104–116.

DE BACKER, A., MARTINEZ, G. T., MACARTHUR, K. E., JONES, L., BÉCHÉ, A., NELLIST, P. D. & VAN AERT, S. (2015). Dose limited reliability of quantitative annular dark field scanning transmission electron microscopy for nano-particle atom-counting. *Ultramicroscopy* **151**, 56–61.

BÉCHÉ, A., GORIS, B., FREITAG, B. & VERBEECK, J. (2016). Development of a fast electromagnetic beam blanker for compressed sensing in scanning transmission electron microscopy. *Applied Physics Letters* **108**.

VAN DEN BROEK, W., REED, B. W., BECHE, A., VELAZCO, A., VERBEECK, J. & KOCH, C. T. (2019). Various Compressed Sensing Setups Evaluated Against Shannon Sampling Under Constraint of Constant Illumination. *IEEE Transactions on Computational Imaging* **5**, 502–514.

BUBAN, J. P., RAMASSE, Q., GIPSON, B., BROWNING, N. D. & STAHLBERG, H. (2010). High-resolution low-dose scanning transmission electron microscopy. *Journal of Electron Microscopy* **59**, 103–112.

BUZHAN, P., DOLGOSHEIN, B., FILATOV, L., ILYIN, A., KANTZEROV, V., KAPLIN, V., KARAKASH, A., KAYUMOV, F., KLEMIN, S., POPOVA, E. & SMIRNOV, S. (2003). Silicon photomultiplier and its possible applications. In *Nuclear Instruments and Methods in Physics Research, Section A: Accelerators, Spectrometers, Detectors and Associated Equipment* vol. 504, pp. 48–52.

DENG, Z., XIE, Q., DUAN, Z. & XIAO, P. (2013). Scintillation event energy measurement via a pulse model based iterative deconvolution method. *Physics in Medicine and Biology* **58**, 7815–7827.

EGERTON, R. F. (2014). Choice of operating voltage for a transmission electron microscope. *Ultramicroscopy* **145**, 85–93.

EGERTON, R. F., LI, P. & MALAC, M. (2004). Radiation damage in the TEM and SEM. *Micron* **35**, 399–409.

FATERMANS, J., VAN AERT, S. & DEN DEKKER, A. J. (2019). The maximum a posteriori probability rule for atom column detection from HAADF STEM images. *Ultramicroscopy* **201**, 81–91.

FINDLAY, S. D. & LEBEAU, J. M. (2013). Detector non-uniformity in scanning transmission electron microscopy. *Ultramicroscopy* **124**, 52–60.

FRANK, J. (2002). Single-Particle Imaging of Macromolecules by Cryo-Electron Microscopy. *Annual Review of Biophysics and Biomolecular Structure* **31**, 303–319.

GLAESER, R. M. (1971). Limitations to significant information in biological electron microscopy as a result of radiation damage. *Journal of Ultrasructure Research* **36**, 466–482.

GNANASEKARAN, K., DE WITH, G. & FRIEDRICH, H. (2018). Quantification and optimization of ADF-STEM image contrast for beam-sensitive materials. *Royal Society Open Science* **5**, 171838.

ISHIKAWA, R., LUPINI, A. R., FINDLAY, S. D. & PENNYCOOK, S. J. (2014). Quantitative annular dark field electron microscopy using single electron signals. *Microscopy and Microanalysis* **20**, 99–110.

JIANG, N. & SPENCE, J. C. H. (2012). On the dose-rate threshold of beam damage in TEM. *Ultramicroscopy* **113**, 77–82.

JOHNSTON-PECK, A. C., DUCHENE, J. S., ROBERTS, A. D., WEI, W. D. & HERZING, A. A. (2016). Dose-rate-dependent damage of cerium dioxide in the scanning transmission electron microscope. *Ultramicroscopy* **170**, 1–9.

JONES, L. (2016). Quantitative ADF STEM: Acquisition, analysis and interpretation. In *IOP Conference Series: Materials Science and Engineering* vol. 109.

JONES, L., VARAMBHIA, A., BEANLAND, R., KEPAPTSOGLOU, D., GRIFFITHS, I., ISHIZUKA, A., AZOUGH, F., FREER, R., ISHIZUKA, K., CHERNS, D., RAMASSE, Q. M., LOZANO-PEREZ, S. & NELLIST, P. D. (2018). Managing dose-, damage- and data-rates in multi-frame spectrum-imaging. *Microscopy* **67**, 98–113.

JONES, L., YANG, H., PENNYCOOK, T. J., MARSHALL, M. S. J., VAN AERT, S., BROWNING, N. D., CASTELL, M. R. & NELLIST, P. D. (2015). Smart Align—a new tool for robust non-rigid registration of scanning microscope data. *Advanced Structural and Chemical Imaging* **1**.

KOVARIK, L., STEVENS, A., LIYU, A. & BROWNING, N. D. (2016). Implementing an accurate and rapid sparse sampling approach for low-dose atomic resolution STEM imaging. *Applied Physics Letters* **109**.

KRAUSE, F. F., SCHOWALTER, M., GRIEB, T., MÜLLER-CASPARY, K., MEHRTENS, T. & ROSENAUER, A. (2016). Effects of instrument imperfections on quantitative scanning transmission electron microscopy. *Ultramicroscopy* **161**, 146–160.

KRIVANEK, O. L., CHISHOLM, M. F., NICOLOSI, V., PENNYCOOK, T. J., CORBIN, G. J., DELLBY, N., MURFITT, M., OWN, C. S., SZILAGYI, Z. S., OXLEY, M. P., PANTELIDES, S. T. & PENNYCOOK, S. J. (2010). Atom-by-atom structural and chemical analysis by annular dark-field electron microscopy. *Nature* **464**, 571–4.

MACARTHUR, K. E., JONES, L. B. & NELLIST, P. D. (2014). How flat is your detector? Non-uniform annular detector sensitivity in STEM quantification. In *Journal of Physics: Conference Series* vol. 522.

MCMULLAN, G., CHEN, S., HENDERSON, R. & FARUQI, A. R. (2009). Detective quantum efficiency of electron area detectors in electron microscopy. *Ultramicroscopy* **109**, 1126–1143.

MITTELBERGER, A., KRAMBERGER, C. & MEYER, J. C. (2018). Software electron counting for low-dose scanning transmission electron microscopy. *Ultramicroscopy* **188**, 1–7.

PAN, M. & CROZIER, P. A. (1993). Quantitative imaging and diffraction of zeolites using a slow-scan CCD camera. *Ultramicroscopy* **52**, 487–498.

PENNYCOOK, T. J., LUPINI, A. R., YANG, H., MURFITT, M. F., JONES, L. & NELLIST, P. D. (2015). Efficient phase contrast imaging in STEM using a pixelated detector. Part 1: Experimental demonstration at atomic resolution. *Ultramicroscopy* **151**, 160–167.

REVOL, J. F. & MANLEY, R. S. J. (1986). Lattice imaging in polyethylene single crystals. *Journal of Materials Science Letters* **5**, 249–251.

S'ARI, M., CATTLE, J., HONDOW, N., BRYDSON, R. & BROWN, A. (2019). Low dose scanning transmission electron microscopy of organic crystals by scanning moiré fringes. *Micron* **120**, 1–9.

SANDERS, T. & DWYER, C. (2020). Inpainting Versus Denoising for Dose Reduction in Scanning-Beam Microscopies. *IEEE Transactions on Image Processing* **29**, 351–359.

SANG, X. & LEBEAU, J. M. (2016). Characterizing the response of a scintillator-based detector to single electrons. *Ultramicroscopy* **161**, 3–9.

SHANNON, C. E. (1949). Communication in the Presence of Noise. *Proceedings of the IRE* **37**, 10–21.

SHIBATA, N., FINDLAY, S. D., KOHNO, Y., SAWADA, H., KONDO, Y. & IKUHARA, Y. (2012). Differential phase-contrast microscopy at atomic resolution. *Nature Physics* **8**, 611–615.

SHIBATA, N., KOHNO, Y., FINDLAY, S. D., SAWADA, H., KONDO, Y. & IKUHARA, Y.

(2010). New area detector for atomic-resolution scanning transmission electron microscopy. *Journal of Electron Microscopy* **59**, 473–479.

STEVENS, A., YANG, H., CARIN, L., ARSLAN, I. & BROWNING, N. D. (2014). The potential for Bayesian compressive sensing to significantly reduce electron dose in high-resolution STEM images. *Microscopy* **63**, 41–51.

UCHIKAWA, Y., GOUHARA, K., YAMADA, S., ITO, T., KODAMA, T. & SARDESHMUKH, P. (1992). Comparative Study of Electron Counting and Conventional Analogue Detection of Secondary Electrons in SEM. *Electron Microsc* **41**, 253–260.

YAMADA, S., ITO, T., GOUHARA, K. & UCHIKAWA, Y. (1991). Electron-count imaging in SEM. *Scanning* **13**, 165–171.

YANG, H., RUTTE, R. N., JONES, L., SIMSON, M., SAGAWA, R., RYLL, H., HUTH, M., PENNYCOOK, T. J., GREEN, M. L. H., SOLTAU, H., KONDO, Y., DAVIS, B. G. & NELLIST, P. D. (2016). Simultaneous atomic-resolution electron ptychography and Z-contrast imaging of light and heavy elements in complex nanostructures. *Nature Communications* **7**.

**Supplementary Information**

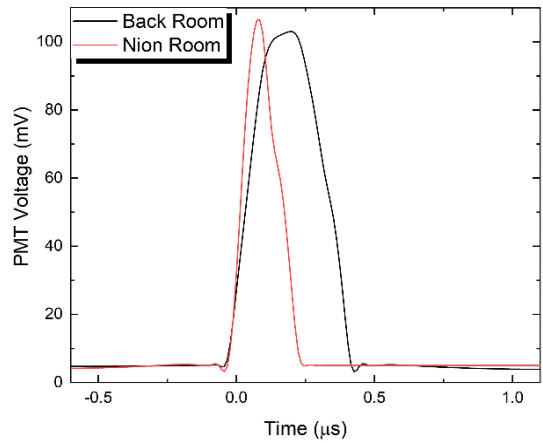

*Figure S1. Comparison of pulses captured directly from the ADF detector (Nion Room), and from the electronics room (Back Room)*

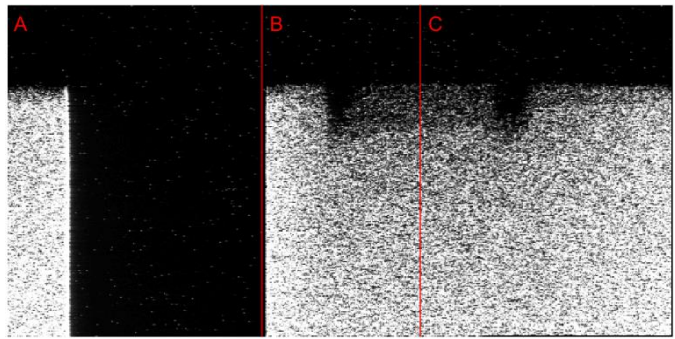

*Figure S2. Untrimmed digital image of the silicon lamella from* Figure 5. *In this untrimmed image the vertical red lines separate the image into three sections labelled A, B, and C. "A" is the image normally presented to the user. "B" is the part of the flyback time where the beam is travelling left, overshooting the original field of view. "C" is where the beam is travelling back to the right the begin the next line of the image.*

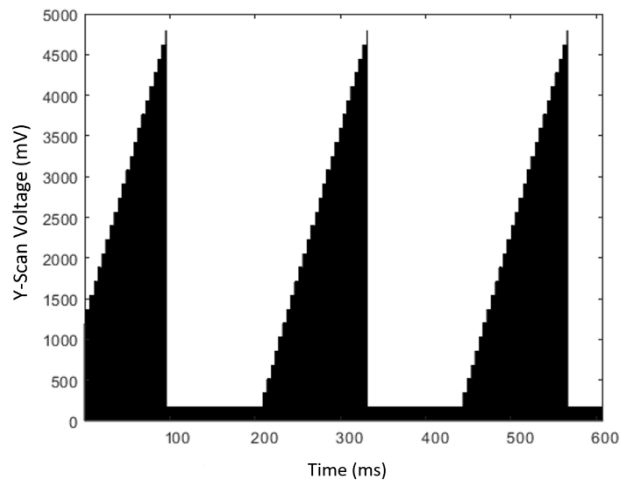

*Figure S3. The sawtooth waveform which drives the y-scan on the Titan microscope. The voltage increases as the scan progresses and is constant between frames. The step-like nature is due to the bit-depth of the oscilloscope used.*